\documentclass{ifacconf}

\usepackage{booktabs}
\usepackage{siunitx}

\usepackage[pdftex]{graphicx}	      
\usepackage{mathtools}
\usepackage{amsmath}
\usepackage[utf8]{inputenc}
\usepackage{array}
\usepackage{wrapfig}
\usepackage{subcaption}  
\usepackage{epstopdf}
\usepackage{commath}
\usepackage{tabularx} %
\usepackage{fancyhdr}
\usepackage{multirow}
\usepackage{comment}
\usepackage{textcomp}

\usepackage{etoolbox}

\AtBeginEnvironment{figure}{\setlength{\captionwidth}{0.8\linewidth}}
\AtBeginEnvironment{table}{\setlength{\captionwidth}{0.8\linewidth}}

\usepackage{natbib}        

\begin{document}
\begin{frontmatter}

\title{Rack Force Estimation for Driving on Uneven Road Surfaces
} 

\thanks[footnoteinfo]{This work was supported by Ford Motor Company under a Ford/U-M Alliance
Project UM0146. \\\textcopyright\hspace{0.4em} 2020 the authors. This work has been accepted to IFAC for publication under a Creative Commons Licence CC-BY-NC-ND.}

\author[First]{Akshay Bhardwaj} 
\author[Second]{Daniel Slavin} 
\author[Second]{John Walsh} 
\author[Third]{James Freudenberg} 
\author[First]{R. Brent Gillespie} 

\address[First]{Mechanical Engineering, University of Michigan, Ann Arbor, MI 48104 USA (e-mail: akshaybh@umich.edu, brentg@umich.edu).}
\address[Second]{Ford Motor Company, Dearborn, MI 48126 USA\\(e-mail: dslavin3@ford.com, jwalsh1@ford.com).}
\address[Third]{Electrical and Computer Engineering, University of Michigan, Ann Arbor, MI 48104 USA (e-mail: jfr@umich.edu).}

\begin{abstract}                

The force transmitted from the front tires and tie rods to the steering rack of a vehicle, called the rack force, significantly influences the torque experienced by a driver at the steering wheel. As a result, estimates of rack force are used in a wide variety of advanced driver assist systems. Existing methods for producing rack force estimates are either susceptible to steering system disturbances or are only applicable for driving on roads with low frequency profile variations such as road slopes. In this paper we present a model that can produce disturbance-free rack force estimates for driving on roads with high frequency profile variations, such as road cleats and potholes, in addition to roads with low frequency profile variations. We validate the estimation accuracy of our model by presenting results from two driving experiments that were performed on test tracks with known low and high frequency road profile variations. We further demonstrate the merits of our model relative to the existing models by comparing the various estimates to rack force measurements obtained using a sensor mounted in the test vehicle.
\end{abstract}


\end{frontmatter}

\section{Introduction}

The torque feedback experienced by a driver at the steering wheel plays an important role in the lateral control of a vehicle. The steering feedback couples the driver's hands and arms to the vehicle and makes the driver aware of the state of the vehicle, the road conditions, and the environment. This enables the driver to plan and choose the driving inputs that would achieve a desired vehicle trajectory and is therefore critical to smooth and controlled vehicle maneuvers (\cite{dornhege2017steering}).

A significant portion of the steering torque feedback comes from the steering rack force, which is defined as the force transmitted from tires to the steering rack of a vehicle through the tie rods. When a driver performs a maneuver, forces and moments are generated at the tire contact patch (as a result of tire-road interaction) that counteract the effort applied by the driver. These forces and moments get transmitted to the steering rack through the tie rods as the rack force that in turn is transmitted as a torque feedback at the steering wheel through the steering pinion and steering column. The tire forces and moments, and hence the rack force, depend on the driver's inputs (such as steering angle and speed) and the road profile variations (such as road slopes, cleats, and potholes).

Apart from the rack force, the steering torque feedback is also influenced by disturbance forces that result from the elements internal to the steering system, for example, suspension asymmetries, steering system friction, wheel rotor imbalance, and brake torque variations (\cite{pick, dornhege2017steering}). Such disturbances must be rejected as they are not useful feedback to the driver and can potentially be dangerous. Likewise, it is also important to attenuate the influence of steering rack force on the torque feedback so that the driver has to apply less effort to counteract the rack force. However, the rack force must not be completely rejected as it is important for the driver to maintain awareness of the state of the vehicle and the road conditions.

To address these requirements, modern electric power steering (EPS) systems utilize estimates of steering rack force (\cite{gruner2008control}) to overlay controlled amounts of torque at the steering wheel to improve the quality of steering feedback. For example, EPS applications such as disturbance rejection controllers (\cite{dornhege2017steering}, \cite{pick}, \cite{blommer2012systems}), lane keeping assist systems, and steer-by-wire position controllers (\cite{fankem2014model}, \cite{nehaoua2012rack}) use the estimates of rack force in their control algorithms. Apart from EPS, rack force estimation is also useful in designing steering feedback in driving simulators, hardware in the loop simulators (\cite{nehaoua2012rack, segawa2006preliminary}), and in performing steering system evaluation and bench-marking (\cite{wang2016epas}). Unfortunately, reliable and durable measurement systems for rack force are expensive. Therefore, estimation of rack force using real time capable models has been of recent interest in the literature (\cite{dornhege2017steering, pick, fankem2014model, blommer2012systems, nehaoua2012rack, wang2016epas}). 


Three real-time capable methods to estimate the rack force exist in the literature. The first method is based on identifying a model for rack force based on experimental data using system identification techniques (\cite{wang2016epas,blommer2012systems}). In \cite{wang2016epas}, several driving experiments were conducted, and a rack force model was identified by characterizing the relationship between measurements of rack force and rack displacement. A disadvantage of system identification based rack force estimation methods is in their limited applicability in maneuvers and vehicle configurations different from the ones for which they are identified. Also, currently these methods do not suggest a way to incorporate road profile variations in the models. The second rack force estimation method that exists in the literature is based on an input observer that uses steering system sensors and a lumped parameter model of the steering system (\cite{dornhege2017steering, fankem2014model, blommer2012systems, nehaoua2012rack}). The rack force estimates produced by this method are applicable in a wide variety of maneuvers and road profile variations and have been validated using measurements from rack force sensors (\cite{dornhege2017steering, fankem2014model, blommer2012systems}). However, the estimates of rack force produced by this method are susceptible to disturbance forces that act within the steering system as shown in \cite{dornhege2017steering}.

The third and final rack force estimation method uses vehicle and tire dynamics models to determine the steering rack force based on sensed driver inputs (\cite{dornhege2017steering,pick,koch2010untersuchungen,segawa2006preliminary}). By using only the driver inputs, this method eliminates the presence of any disturbance forces in the estimated rack force (\cite{dornhege2017steering}). However, this method currently ignores the presence of road profile variations and only considers the driver’s inputs when estimating the rack force. In our work, we have attempted to fill this gap by extending the vehicle and tire dynamics based rack force estimation method to incorporate road profile inputs (that is, the dimensions of road unevenness).

In a previous paper, we presented two models, a 2DOF model and a 3DOF model, to estimate rack force (\cite{bhardwaj2019estimating}) due to road profile variations. 
These models could however only capture the effect of lateral road slopes (road banks), and longitudinal road slopes (road grades) on the rack force which are generally categorized as low frequency road profile variations ($<8$Hz) (\cite{pacejka2005tire}). In this paper, we extend our previous work and present a new model that we call RR (Rigid Ring) Model, that can also estimate rack force due to higher frequency road profile variations such as oblique cleats, curbs and potholes ($8$Hz$-80$Hz) (\cite{schmeitz2004semi}, \cite{zegelaar1996plane}). 
To validate the estimation accuracy of the developed RR Model, we present the results from two driving experiments and compare the estimates produced by the RR Model in these experiments with those produced by the rack force sensor mounted in the vehicle, the 3DOF model developed in our previous paper, and a vehicle dynamics model existing in the literature that does not account for road profile variations. 

This paper is structured as follows. In Section 2 we present the modeling framework and present the details of the RR Model followed by the details of the experimental setup in Section 3 that was used to validate the estimation accuracy of the model. Section 4 presents the results of the driving experiments on varying road slopes and on road cleats of known dimensions. Section 5 presents the conclusions of the paper.

\section{Modeling}\label{mod_sec}

\begin{figure}[ht]
  \vspace{0 cm }
\begin{center}
 \includegraphics[width=0.48\textwidth]{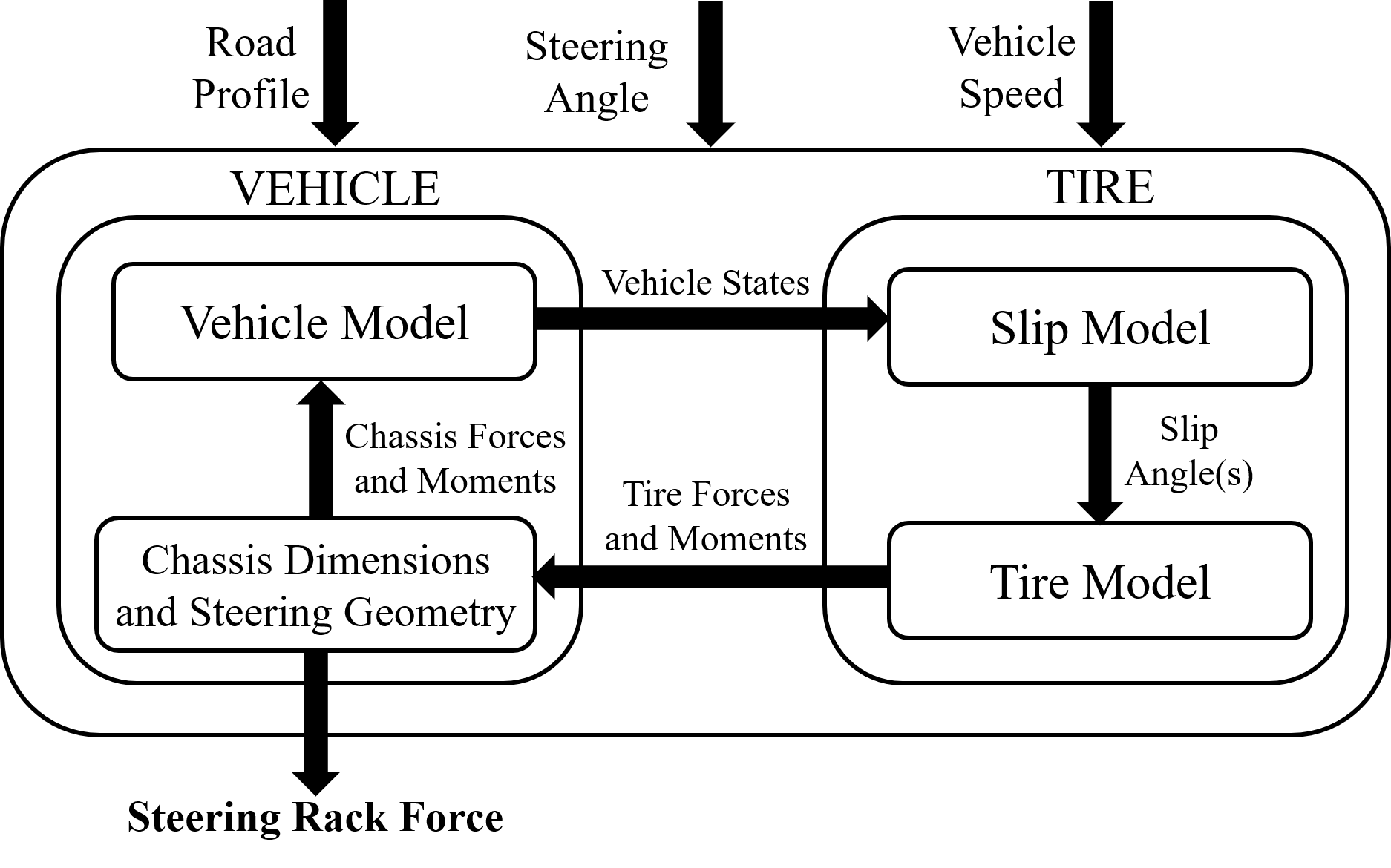}
 \end{center}
 \vspace{-.4 cm }
  \caption{The estimates of steering rack force were produced by a communicating vehicle model and tire model given road profile, steering angle, and vehicle speed. }
  \label{Fig:overall3}
 \end{figure}

Figure \ref{Fig:overall3} describes the overall structure of a dynamic model that can be used to estimate the steering rack force. Three inputs were required in the model; namely, steering angle, vehicle speed, and road profile inputs (which include lateral road slope or longitudinal road slope or cleat/pothole dimensions). 

Using these inputs, a dynamic model of the vehicle was developed to generate the vehicle states (lateral speed, yaw rate, roll angle, etc.) which were fed into a tire slip model that determines the slip angles of the tire. Normal tire forces were computed using the road profile inputs. Normal tire forces along with the slip angles, were used to compute the tire forces and moments which were again used as inputs to the vehicle model to generate the states for the next time interval. During this process, the aligning moments resulting from the tire model were used to determine the steering rack force (using information of the transmission mechanics of tire moments to rack force).

\subsection{Vehicle Model}\label{2dofmodel}

A 2 DOF bicycle model was used as the vehicle model with the RR Model. A detailed derivation of this vehicle model can be found in \cite{bhardwaj2019estimating}. Here we simply provide the equations of motion that produce the vehicle states that are used in the tire model discussed in the next subsection.

Consider a vehicle of mass $m$ and yaw inertia $I$ driving with steering angle $\delta$, and speed $u$. Let the lateral slope of the road be $\theta$, the vehicle yaw angle be $\psi$, and let the forces on the vehicle's front and rear tires in the longitudinal and lateral directions respectively be denoted by $F_{xf},F_{xr}$ and $F_{yf},F_{yr}$. Then assuming steering angle $\delta$ is small, the two degrees of freedom, namely, lateral speed $v$ and yaw rate $\dot{\psi}$ for driving on lateral slope are governed by the following differential equations
\begin{gather}\label{dyn_lat1}
\begin{aligned}
m\dot{v}+ mu\dot\psi+mg\sin\theta = F_{yf}+F_{yr}\\
{I}\ddot{\psi} =l_fF_{yf}-l_rF_{yr},
\end{aligned}
\end{gather}
Where $l_f$ and $l_r$ are the distances from the vehicle's center of gravity to the front and rear axles, respectively.

For driving on a longitudinal slope, the lateral speed and yaw rate are simply governed by the following equations
\begin{gather}\label{dyn_long}
\begin{aligned}
m\dot{v}+ mu\dot\psi =F_{yf}+F_{yr}\\
{I}\ddot{\psi} =l_fF_{yf}-l_rF_{yr}
\end{aligned}
\end{gather}

\subsection{Tire Model}

Longitudinal and lateral slopes are categorized as lower frequency (or longer wavelength) road profile variations. For such variations, the profile (or geometry) of the road surface can be used directly as an input to the vehicle and tire models. However, for higher frequency road profile variations, such as road cleats and potholes, the first step is to reconstruct the profile of the road (\cite{zegelaar1996plane}). This is because while driving on a cleat or a pothole, the tires deform significantly due to their elastic properties. And because of their deformation and geometry, the road profile that the tires experience, called the ``effective road profile", is significantly different from the actual profile of the road on which the tires are traversing (\cite{schmeitz2004semi}). One often says that the tires filter (or smoothen) the sharp edges of road cleats and potholes (\cite{zegelaar1996plane}). 



As mentioned in \cite{zegelaar1996plane}, an effective road profile can be generated using a ``tire enveloping model". In this paper, we used a semi-empirical three dimensional tire enveloping model introduced in \cite{schmeitz2004semi} to determine the effective road profile. This model produces an effective road profile in the form of three road geometry parameters: effective tire height $w$, effective tire lateral slope $\beta_x$, and effective tire longitudinal slope $\beta_y$. Inputs to the tire enveloping model are the dimensions of the cleat (length, height, width, and angle), and the location of the cleats. 
A detailed description of this tire enveloping model can be found in \cite{schmeitz2004semi} (Chapter 4). The purpose of this section is to describe the simplified Rigid Ring tire model that was used to obtain the tire forces and moments using the effective road profile obtained from the enveloping model. 

\begin{figure}[ht]
  \vspace{0 cm }
\begin{center}
 \includegraphics[width=0.325\textwidth]{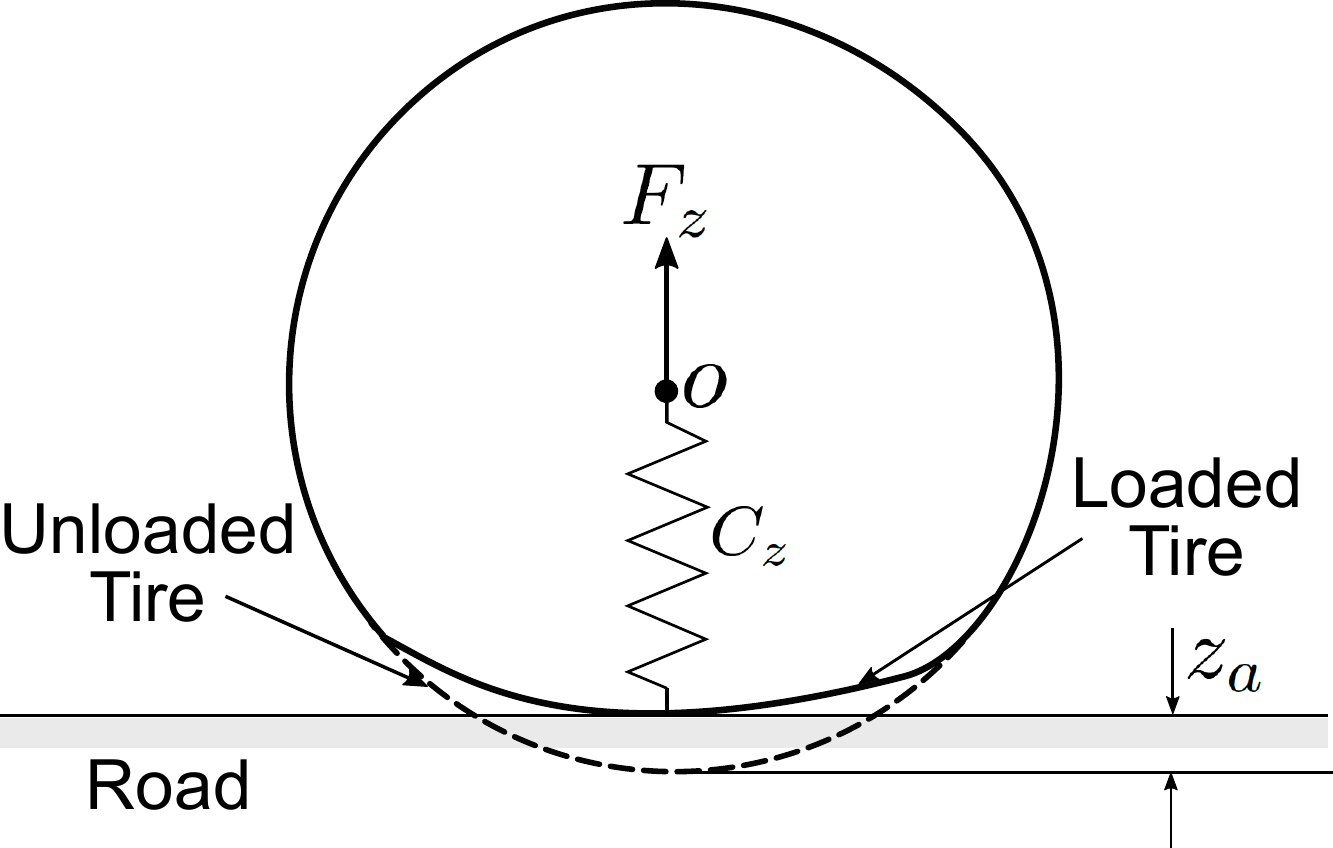}
 \end{center}
 \vspace{-.4 cm }
  \caption{A tire rolling over a flat road profile experiences a vertical tire deflection $z_a$ that is roughly proportional to the normal load $F_z$ acting on the tire.}
  \label{Fig:flat}
 \end{figure}

Consider the case of driving on a flat road with no unevenness as shown in Fig. \ref{Fig:flat}. Suppose the tire has a vertical stiffness $C_z$ and suppose that under a normal force $F_z$ (acting due to vehicle's weight), the tire displaces vertically by a distance $z_a$. Then, the normal force $F_z$ for this tire can be given by
\begin{gather}\label{nom2}
F_z = C_{z}z_a
\end{gather}

Also, by applying a vertical force balance on a vehicle traveling on a road with slope $\theta$ (lateral or longitudinal), the normal force $F_z$ can be written as
\begin{gather}\label{nom}
F_{z} =
	\begin{dcases}
		\frac{mgl_r\cos\theta}{2(l_f+l_r)} & \quad \mbox{for the front tires} \\
		\frac{mgl_f\cos\theta}{2(l_f+l_r)} & \quad \mbox{for the rear tires}
	\end{dcases}
\end{gather}

Then using equations (\ref{nom2}) and (\ref{nom}), $z_a$ can be determined as follows
\begin{gather}\label{za}
z_a =
	\begin{dcases}
		\frac{mgl_r\cos\theta}{2C_{z}(l_f+l_r)} & \quad \mbox{for the front tires} \\
		\frac{mgl_f\cos\theta}{2C_{z}(l_f+l_r)} & \quad \mbox{for the rear tires}
	\end{dcases}
\end{gather}

\begin{figure}[h!]
  \vspace{0 cm }
\begin{center}
 \includegraphics[width=0.48\textwidth]{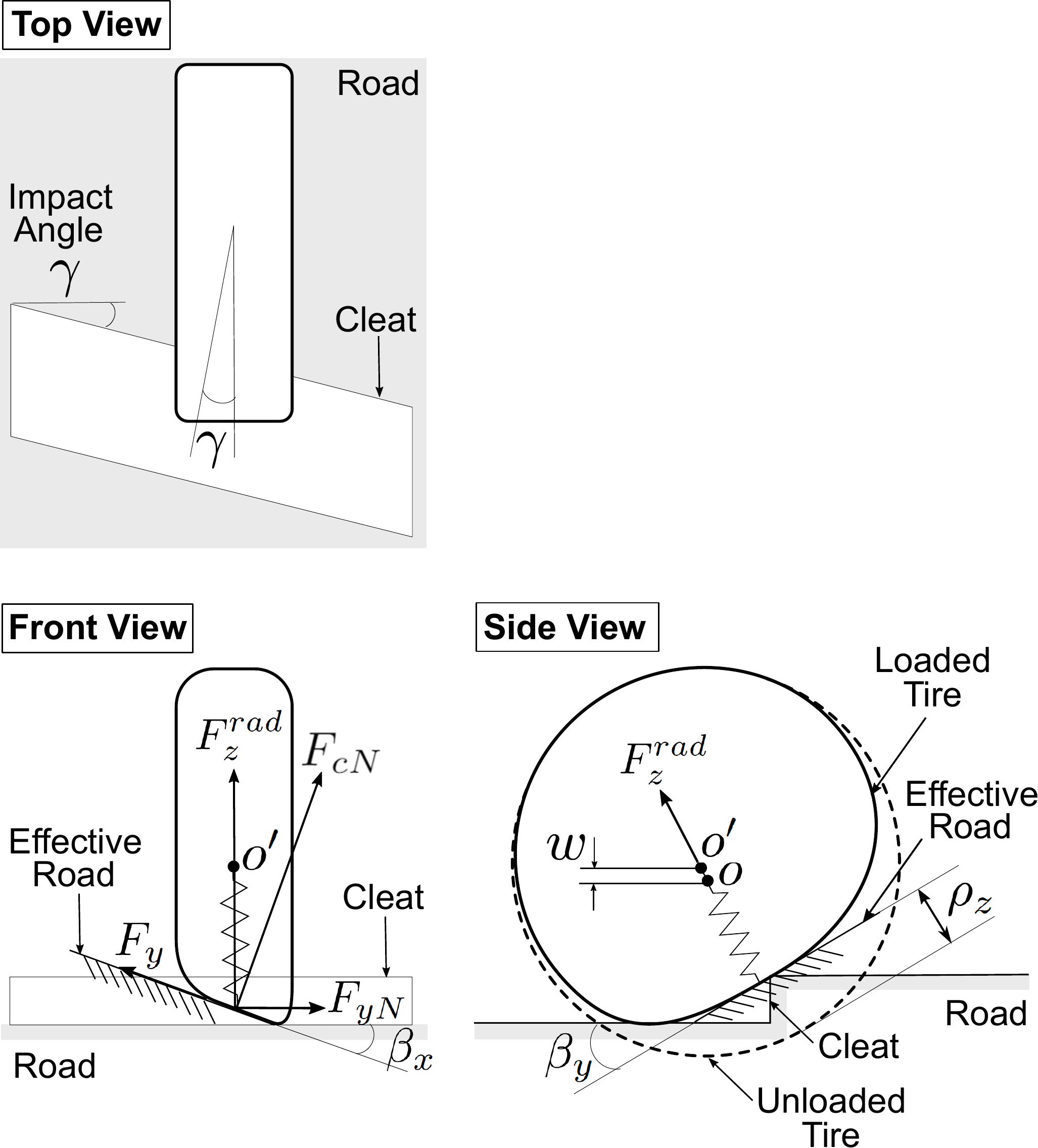}
 \end{center}
 \vspace{-.4 cm }
  \caption{A tire rolling over a high frequency road unevenness such as an oblique road cleat experiences a road profile different from the actual profile of the road. Three orthogonal views demonstrate how a road with an oblique cleat can be transformed into an effective road profile with an effective lateral slope $\beta_x$, an effective longitudinal slope $\beta_y$, and an effective tire height $w$. The forces acting on the tire on an effective road profile are also shown.}
  \label{Fig:cleat}
 \end{figure}

Now consider the case of driving on an arbitrary high frequency road unevenness such as an oblique cleat. Fig. \ref{Fig:cleat} shows three orthogonal views of a tire hitting a cleat at an impact angle $\gamma$. We will use this figure to develop a model for tire lateral force $F_y$ that is used to estimate the associated tire aligning moment and rack force.

Looking at the side view of Fig. \ref{Fig:cleat}, we see that when a tire hits an oblique cleat, it displaces upwards and the tire center $o$ moves to $o'$. The vertical distance between $o$ and $o'$ is denoted by $w$ and the overall displacement of the loaded tire on the cleat with respect to the unloaded tire is denoted by $\rho_z$. Moreover, the deformation of the tire converts the ``step" profile of the cleat into an effective longitudinal tire slope $\beta_y$. As mentioned previously, both $w$ and $\beta_y$ can be obtained using semi-empirical tire enveloping models. Displacement $w$ and angle $\beta_y$ along with $z_a$ obtained in Equation (\ref{za}) can be used to determine the radial deflection of the tire $\rho_z$ as follows
\begin{gather}\label{rho_z}
    \rho_z = ({w-z_a})\cos\beta_y.
\end{gather}

Now, looking at the front view of Fig. \ref{Fig:cleat}, we see that along with experiencing a longitudinal tire slope $\beta_y$, the tire also experiences a lateral slope $\beta_x$ due to the obliqueness of the cleat. Angle $\beta_x$ along with tire deflection $\rho_z$ (Equation \ref{rho_z}) can be used to estimate the radial tire force using the following semi-empirical equation (\cite{schmeitz2004semi})
\begin{gather}\label{rad}
    F_z^{rad} = q_{Fz1}(1+q_{Fz3}(\beta_x)^2)\rho_z+q_{Fz2}\rho_z^2,
\end{gather}

where $q_{Fz1}$, $q_{Fz2}$, and $q_{Fz3}$ are tire parameters that can be determined using empirical tests. 

When a tire is loaded against a cambered road surface, a side force $F_{yN}$ is developed on it due to non-symmetric deformation of the tire sidewalls. This side force is called the ``non-lagging" lateral force. The effective lateral slope $\beta_x$, and the radial tire force $F_z^{rad}$ (obtained in equation (\ref{rad})) can be used to determine the non-lagging force $F_{yN}$ using the following empirical model
\begin{gather}\label{lag}
\begin{aligned}
    F_{yN} = (D_{N}\sin(C_N\arctan (B_N \beta_x)))\cos \beta_x -\\ F_z^{rad}\sin \beta_x,
    \end{aligned}
\end{gather}

where the coefficients $D_{N}$, $C_{N}$, and $B_{N}$ are functions of the radial tire force $F_z^{rad}$ and tire normal force $F_z$.

Looking at the front view of Fig. \ref{Fig:cleat}, the non-lagging tire force $F_{yN}$ and the radial tire force $F_z^{rad}$ can then be used to estimate the contact patch force $F_{cN}$ acting normal to the tire contact patch as follows
\begin{gather}\label{fz}
    F_{cN} = \frac{1}{\cos \beta_x}(F_z^{rad}+F_{yN}\sin\beta_x)
\end{gather}

Finally, using the Rigid Ring Tire Model, the tire lateral force $F_y$ can be estimated using the following expression (\cite{schmeitz2004semi})
\begin{gather}\label{forc}
\begin{aligned}
F_{y} = D_y \sin(C_y \arctan\lbrace B_y\alpha_y - E_y(B_y\alpha_y- \\ \arctan(B_y\alpha_y))\rbrace)+S_{Vy},
\end{aligned}
\end{gather}

and the resulting aligning moment $M_z$ acting on a single tire can be estimated using
\begin{gather}\label{al_3}
    M_{z} = -tF_{y}+D_r\cos(\arctan{B_r\alpha_r}),
\end{gather}

where the pneumatic trail $t$ is given by
\begin{gather*}
t = D_t \cos(C_t \arctan\lbrace B_t\alpha_t - E_t(B_t\alpha_t-\arctan(B_t\alpha_t))\rbrace),
\end{gather*}

and the slip angles ($\alpha_y$, $\alpha_t$, and $\alpha_r$) are given by
\begin{gather*}
\alpha_y = S_{Hy}+ \tan{\alpha}, \hspace{2em} 
\alpha_t = S_{Ht}+\tan{\alpha}, \hspace{2em} 
\alpha_r = \tan{\alpha}.
\end{gather*}

The coefficients: $B_y$, $B_r$, $B_t$, $C_y$, $C_t$, $D_y$, $D_r$, $D_t$, $E_y$, $E_t$, $S_{Hy}$, and $S_{Ht}$ are either constants or are functions of slip angles ($\alpha_y$, $\alpha_t$, and $\alpha_r$), tire normal force $F_z$, and contact patch normal force $F_{cN}$ (\cite{schmeitz2004semi}).

The slip angle $\alpha$ can be obtained using the vehicle states obtained in Section \ref{2dofmodel} using the following expression (\cite{ulsoy2012automotive})
\begin{gather}\label{slip}
\alpha = \arctan\left(\frac{v+l_f\dot{\psi}}{u}\right)-\delta,
\end{gather}

Using equation (\ref{al_3}), we can find the aligning moment for the front left tire $M_{z1}$ and for the front right tire $M_{z2}$. The steering rack force is then simply given by
\begin{gather}\label{fr}
F_R = i_p(M_{z1}+M_{z2}),
\end{gather}

where the constant ratio $i_p$ defines the tire moment to rack force transmission ratio for a vehicle.

\subsection{Model Assembly}

Referring back to Fig. \ref{Fig:overall3}, the vehicle states (lateral speed $v$ and yaw rate $\dot{\psi}$) are produced by the vehicle model represented by equations (\ref{dyn_lat1}) and (\ref{dyn_long}). The vehicle states are then used to find the tire slip angle $\alpha$ using equation (\ref{slip}). The vehicle's weight and road profile inputs are used to compute the tire contact patch normal force using equation (\ref{fz}). Tire slip angles and normal forces are used to find tire lateral forces and aligning moments using equations (\ref{forc}) and (\ref{al_3}). The tire forces are fed back in the vehicle model in equation (\ref{dyn_lat1}) and equation (\ref{dyn_long}) to generate the vehicle states for the next time instant. During this process, the rack force for each time instant is obtained using equations (\ref{al_3}) and (\ref{fr}). 

\section{Methods}

To validate the performance of the RR Model for producing estimates of rack force, the following driving experiments were performed:
\begin{enumerate}
    \item \textit{Experiment 1: Driving on a crowned road}\\
    This experiment was performed on a crowned road with a large lateral slope of about  $11^{\circ}$ on the two sides. The vehicle was driven from one side of the road crown to the other side.
    \item \textit{Experiment 2: Driving on a road with cleats}\\
    This experiment was performed on a road with thirteen 4 cm long metal cleats of known heights: the first four cleats were 1 cm tall, the next five cleats were 2 cm tall, and the rest of the cleats were 3 cm tall. The steering angle was intentionally varied between approximately $-50^{\circ}$ and $50^{\circ}$ so that the vehicle impacted the cleats at an angle (roughly equal to the steering angle) to test the performance of the models when driving on arbitrarily uneven roads.
\end{enumerate}

The experiments were conducted on a Lincoln MKX vehicle equipped with Pirelli AS tires. Tire model specific parameters used in the RR Model were taken from \cite{pacejka2005tire} and \cite{schmeitz2004semi}. Other vehicle specific parameters can be found in \cite{bhardwaj2019estimating}. The model-based estimates of rack force were compared to measurements from strain gauges installed on the tie rods of the test vehicle, to the estimates of the 3DOF model developed in our previous work (\cite{bhardwaj2019estimating}), and also to the estimates of a vehicle dynamics model existing in the literature (\cite{dornhege2017steering}). We use the label 2DOF-FR (where FR stands for `flat road') to refer to the existing model since this model uses a 2DOF vehicle model and ignores road profile variations during rack force estimation. 

The steering angle and the longitudinal velocity were measured using steering angle and tire speed sensors respectively. During the driving tests, a rapid control prototyping platform (dSPACE MicroAutoBox) was used to link sensed steering angle and vehicle speed signals with an online simulation of the dynamic models (integrated in real-time Simulink), using CAN-bus communications at 250 Hz. The road slopes (longitudinal and lateral) were measured using slope measurements from a high fidelity IMU (OXTS RT3003 v2) installed in the vehicle that transmitted signals at 100 Hz. The cleat dimensions (height, width and length) were physically measured on the track where the tests were performed.

\section{Results and Discussion}\label{sec:results}

\begin{figure}
	\centering
	\begin{minipage}[b]{.48\textwidth}
		\includegraphics[width=\textwidth]{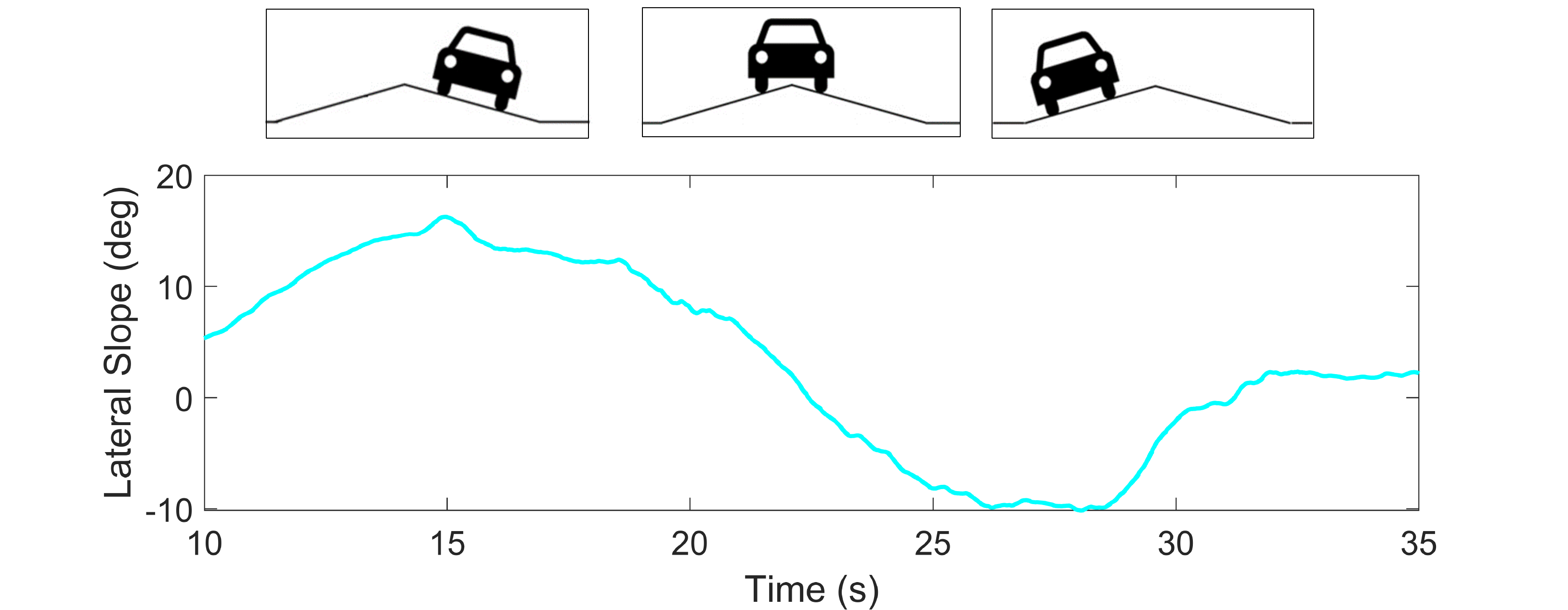}
 \subcaption{Input: Lateral slope of the road}
    \label{1a}
\end{minipage}
	\begin{minipage}[b]{.48\textwidth}
		\includegraphics[width=\textwidth]{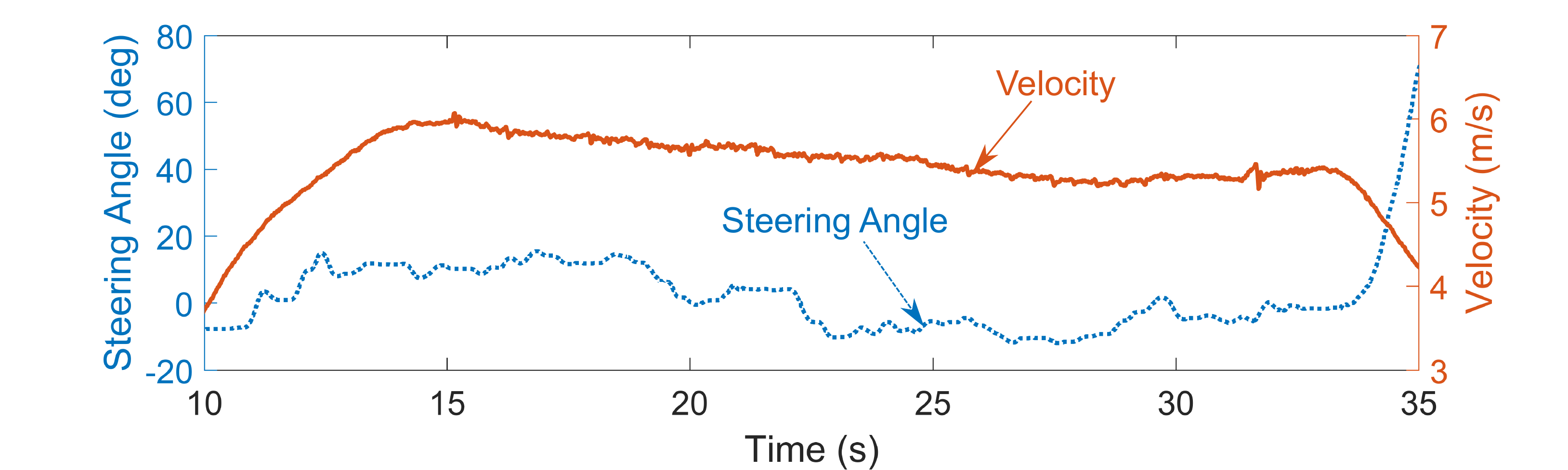}
 \subcaption{Inputs: Steering angle and vehicle speed}
    \label{1b}
\end{minipage}
\begin{minipage}[b]{.48\textwidth}
		\includegraphics[width=\textwidth]{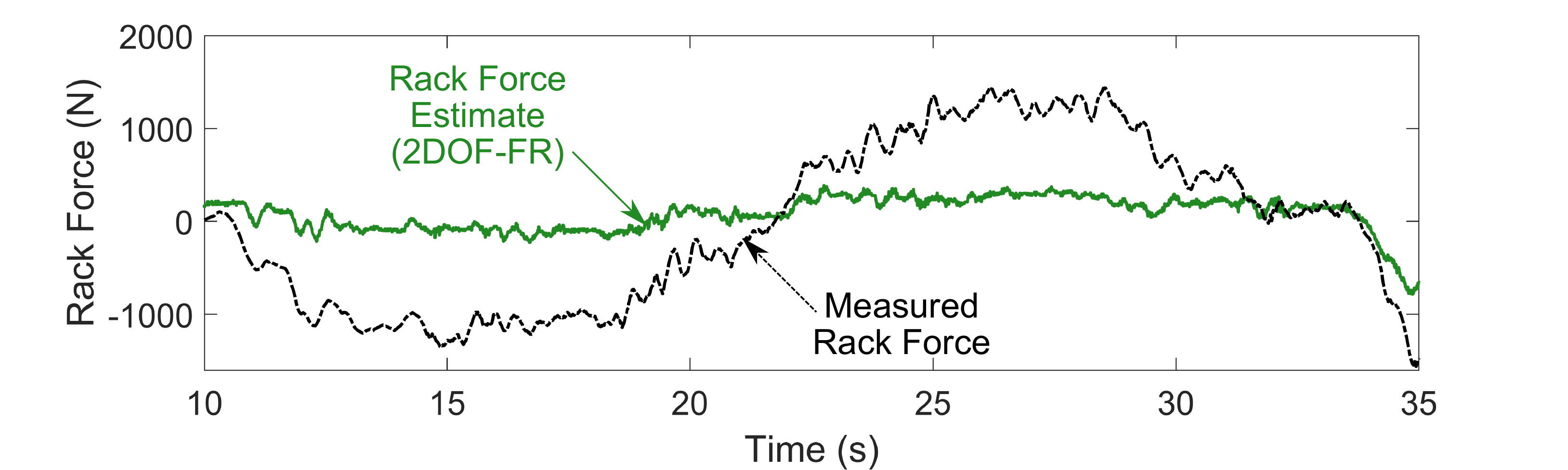}
 \subcaption{Rack force estimated with the 2DOF-FR Model and measured using sensor}
    \label{1c}
\end{minipage}
\begin{minipage}[b]{.48\textwidth}
		\includegraphics[width=\textwidth]{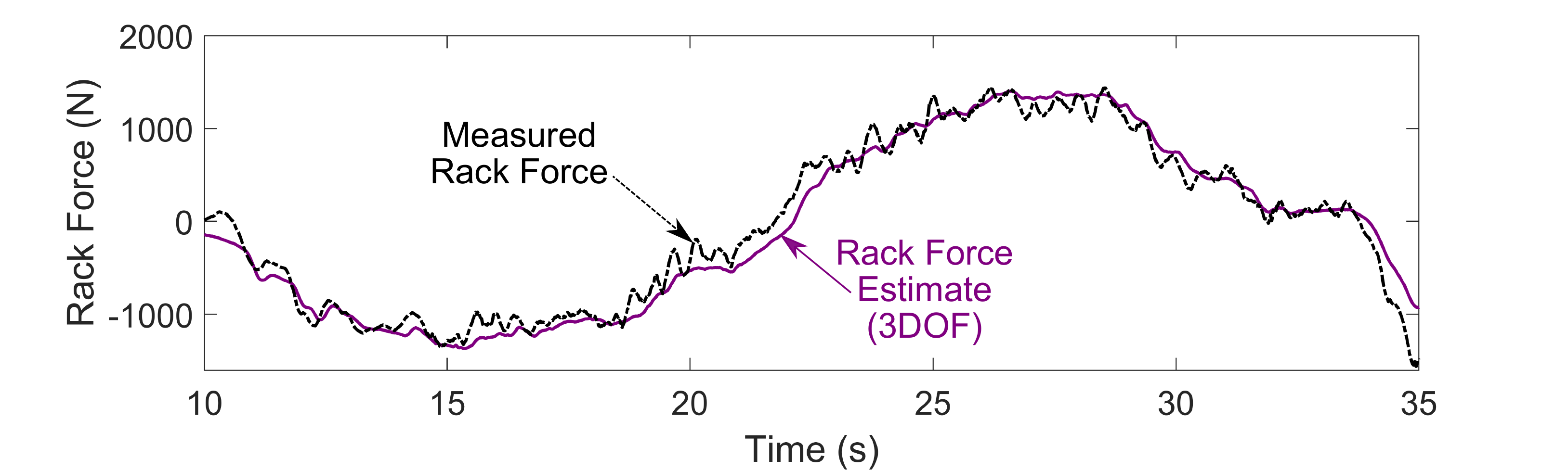}
\subcaption{Rack force estimated with 3DOF Model and measured using sensor}
    \label{1d}
\end{minipage}
\begin{minipage}[b]{.48\textwidth}
		\includegraphics[width=\textwidth]{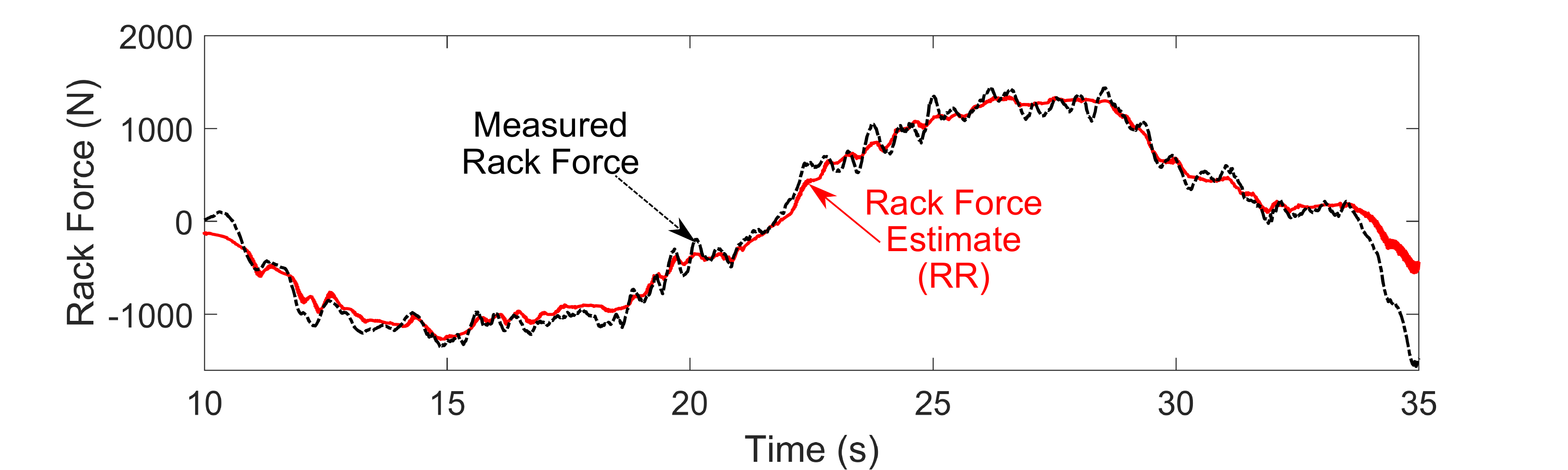}
\subcaption{Rack force estimated with RR Model and measured using sensor}
    \label{1e}
\end{minipage}
	\caption{Experiment 1 performed on a crowned road. (a) Input Lateral Slope. (b) Input steering angle and input velocity (approximately constant at 5.5 m/s). (c) Rack Force estimated using the 2DOF-FR model and measured using sensor. (d) Rack Force estimated using the 3DOF model and measured using sensor. (e) Rack Force estimated using the RR model and measured using sensor.} 
		\label{exp1}
\end{figure}

The differences between the 2DOF-FR Model, the 3DOF Model, and the RR Model were apparent in the comparison of their estimation errors as shown in Table \ref{table:1}.

\begin{table}[h!]
 \setlength\extrarowheight{2pt}
\centering
\caption{Mean absolute estimation errors (N) for the two experiments}
 \begin{tabular}{||c c c||} 
 \hline
 Model/Experiment & Experiment 1 & Experiment 2 \\ [0.5ex] 
  \hline\hline
 2DOF-FR (Flat Road) & 645.14 & 229.36 \\
 3DOF & 108.23 & 193.60 \\
 RR (Rigid Ring) & 101.14 & 136.89 \\
  \hline
\end{tabular}
\label{table:1}
\end{table}

For driving on a crowned road with large slope variation in Experiment 1, we found that the rack force estimates produced by both the 3DOF model and the RR model agreed well with the sensor measurements (see Fig. \ref{exp1}). As expected, the estimation accuracy of the 2DOF-FR Model was lower than both the 3DOF model and the RR model because the 2DOF-FR did not incorporate road slopes in estimation. Interestingly, despite the differences in the tire models used in the 3DOF Model and the RR Model, the estimation performance of the two models were only marginally different from each other. 

\begin{figure}
	\centering
	\begin{minipage}[b]{.48\textwidth}
		\includegraphics[width=\textwidth]{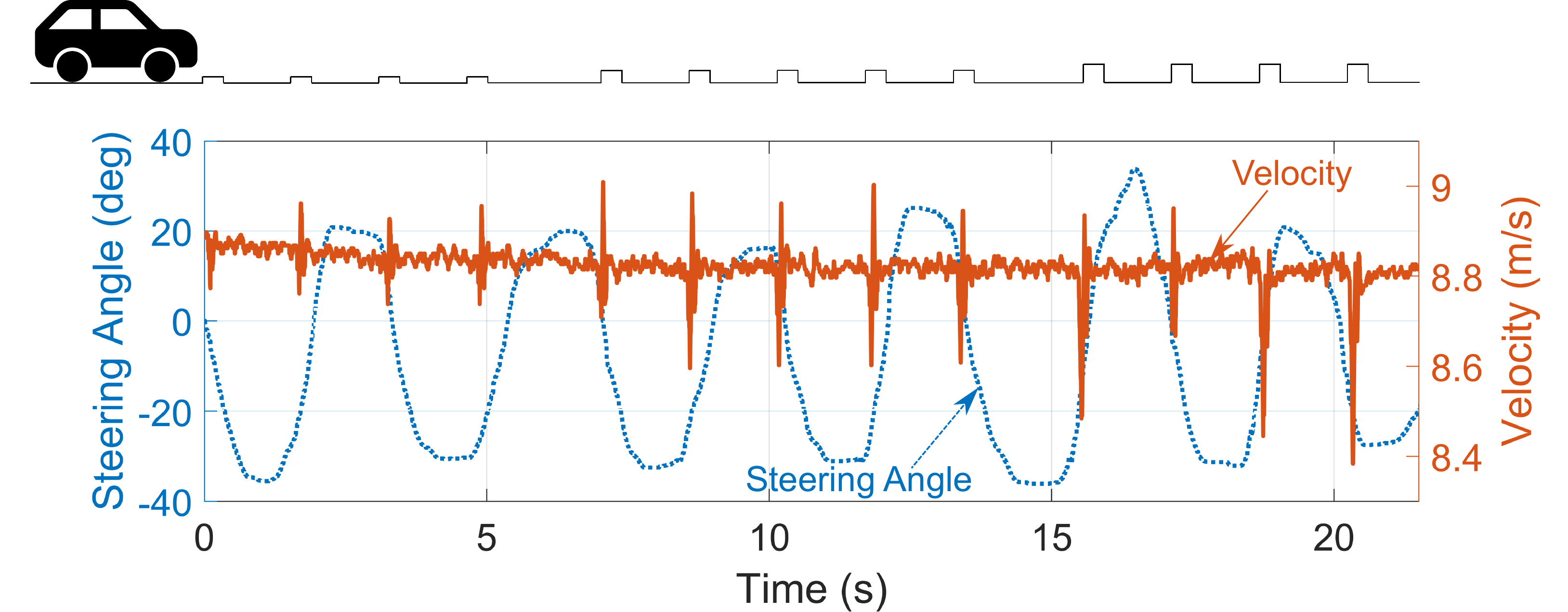}
 \subcaption{Inputs: Steering angle and vehicle speed}
    \label{3a}
\end{minipage}
\begin{minipage}[b]{.48\textwidth}
		\includegraphics[width=\textwidth]{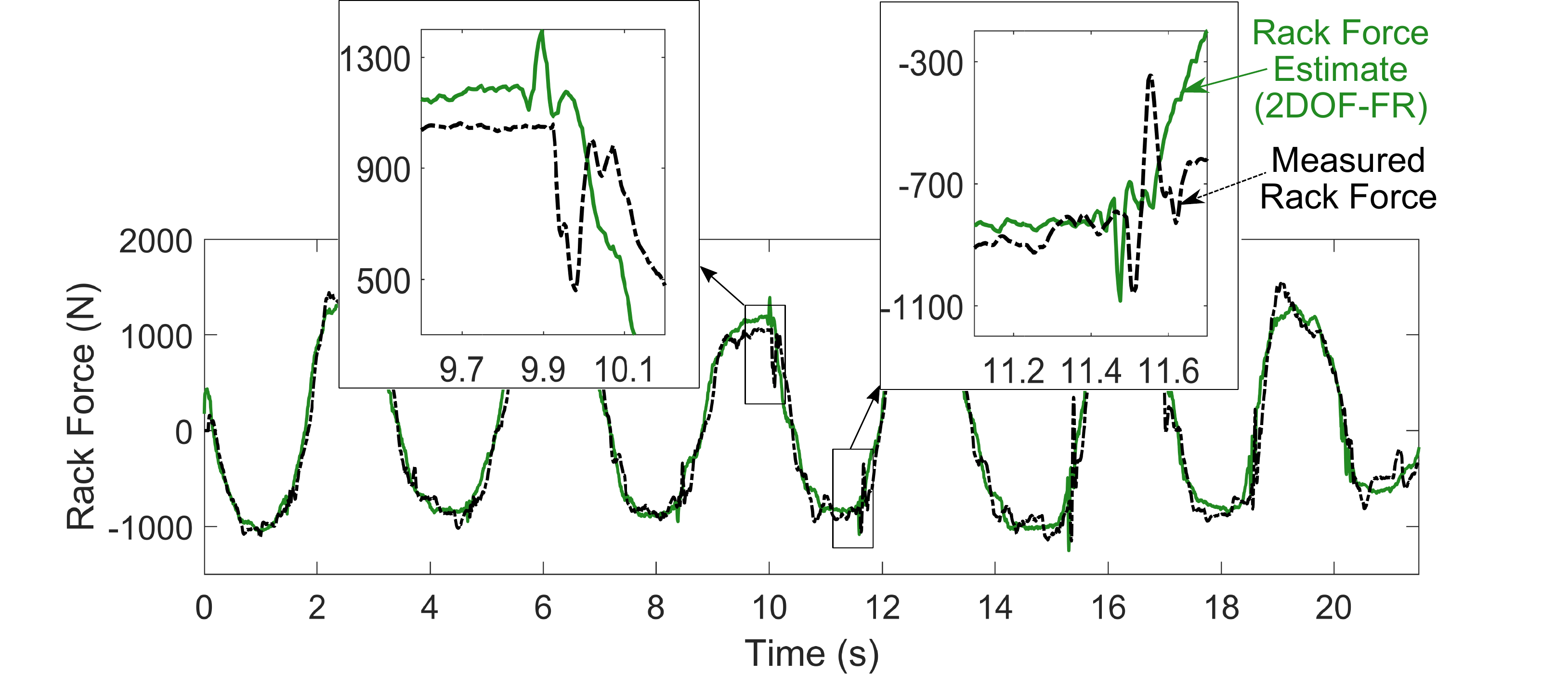}
 \subcaption{Rack force estimated with 2DOF-FR Model and measured using sensor}
    \label{3b}
\end{minipage}
\begin{minipage}[b]{.48\textwidth}
		\includegraphics[width=\textwidth]{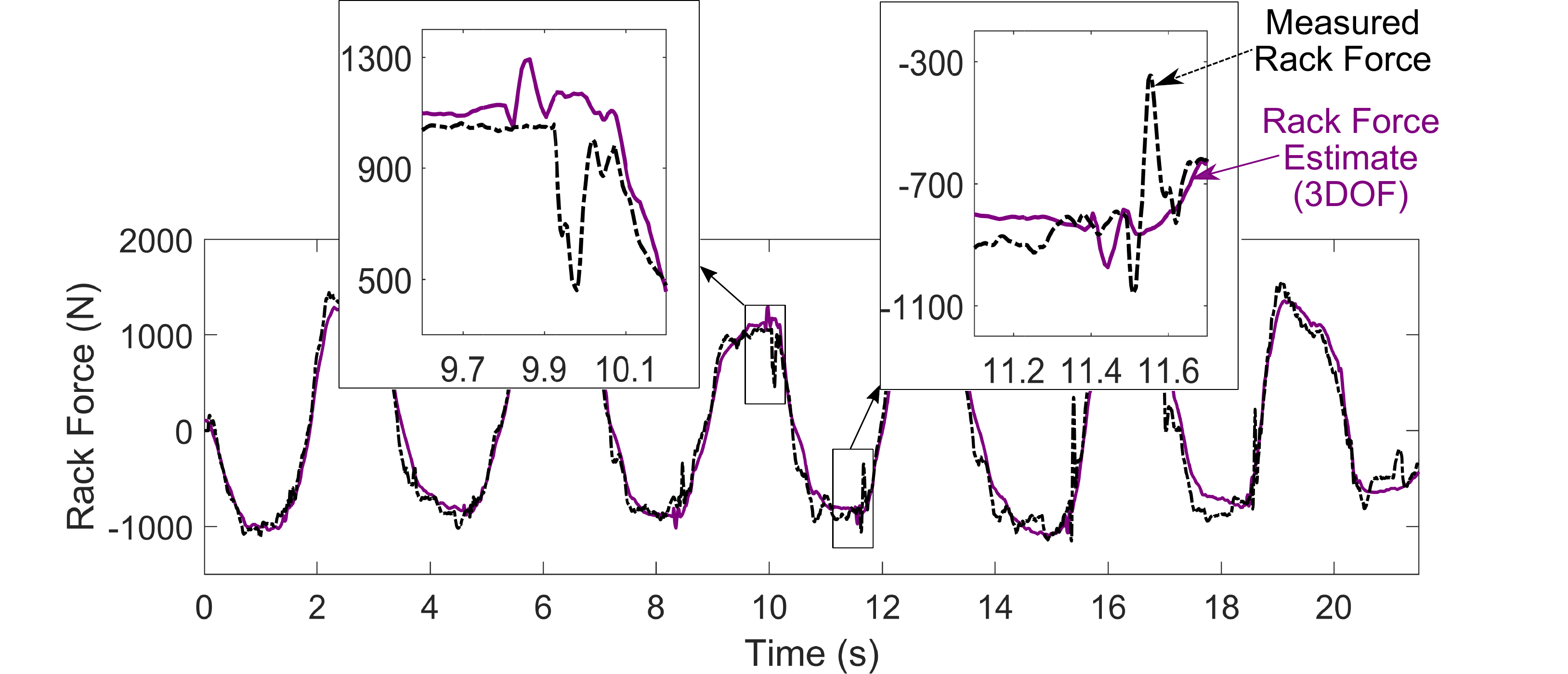}
\subcaption{Rack force estimated with 3DOF Model and measured using sensor}
    \label{3c}
\end{minipage}
\begin{minipage}[b]{.48\textwidth}
		\includegraphics[width=\textwidth]{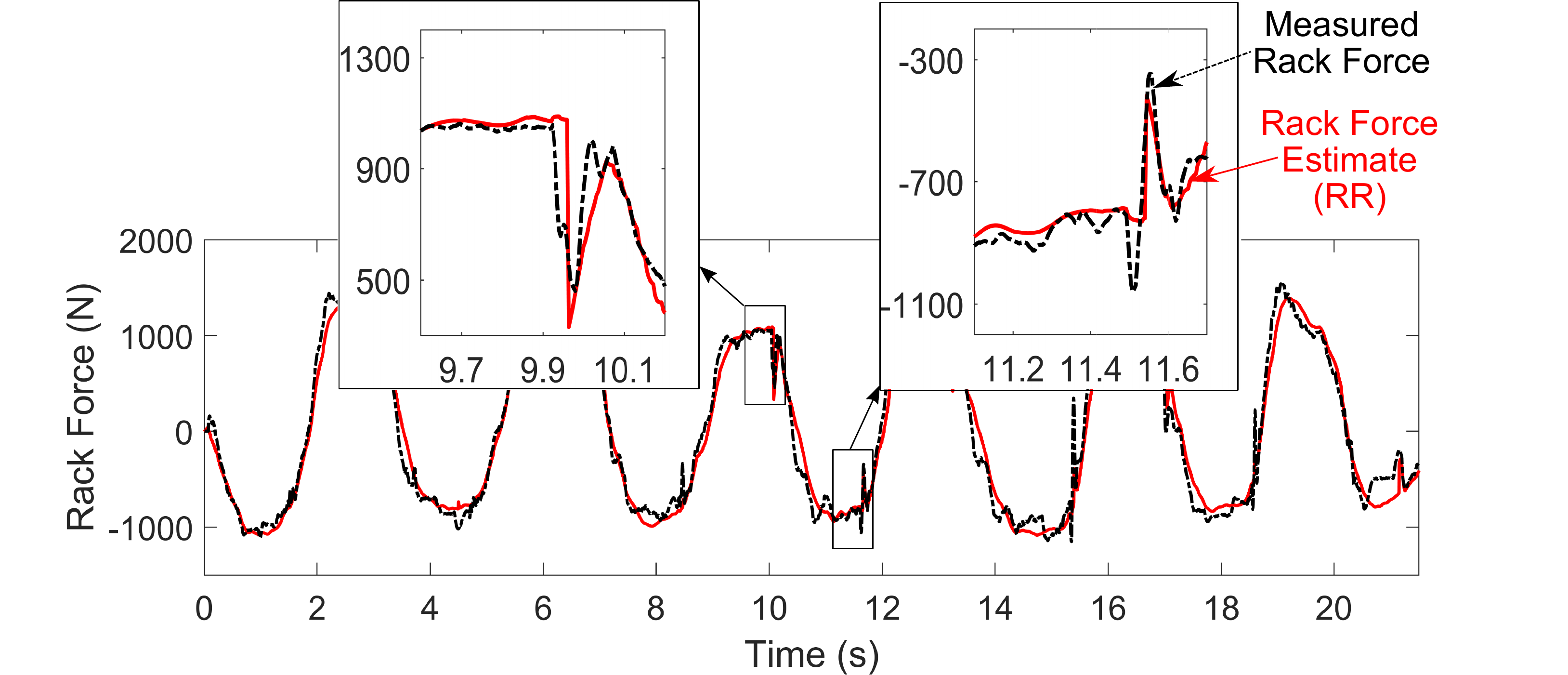}
\subcaption{Rack force estimated with RR Model and measured using sensor}
    \label{3d}
\end{minipage}
	\caption{Experiment 2 performed a road with cleats. (a) Inputs: steering angle (representing a slalom maneuver) and velocity (approximately constant at 8.8m/s). (b) Rack Force estimated using the 2DOF-FR model and measured using sensor. (c) Rack Force estimated using the 3DOF model and measured using sensor. (d) Rack Force estimated using the RR model and measured using sensor.} 
		\label{exp3}
\end{figure}

While driving on road with cleats in Experiment 2, we saw that the RR model showed the highest accuracy out of all the models as shown in Table \ref{table:1}. This was especially apparent at the instances when the tires hit the cleats (as demonstrated by the insets in Fig. \ref{exp3}) indicating that the RR model could successfully account for high frequency road profile variations in the estimation of rack force.

The results from the two experiments showed that both the 3DOF model and the RR Model performed better than the 2DOF-FR model, and could therefore estimate the rack force accurately due to road profile variations (which were ignored in the 2DOF-FR model). 
Moreover, Experiment 1 showed that even though the RR Model used a tire model of higher complexity than the tire model used in the 3DOF model, the performance of the two models were similar for driving on road slopes. On the other hand, in Experiment 2 when the tires hit the oblique cleats, the RR Model demonstrated a higher estimation accuracy as only the Rigid Ring tire model had the ability to capture the effect of tire deformation due to high frequency road profile variations in the estimation of tire forces and moments.

\section{Conclusion}

In this paper, we presented a model that can estimate rack force for driving on uneven roads. We validated the performance of our model by performing driving experiments on two test tracks with known low and high frequency profile variations. Furthermore, we compared the performance of our model with the performance of two existing rack force estimators. We found that our model supports better estimation performance than the existing models for driving on roads with high frequency profile variations, such as road cleats and potholes. However, for low frequency road profile variations, such as road slopes, the estimation accuracy of our model was almost similar to the accuracy of one of the existing models.
In the future, we would like to estimate the effect of tire model, independent of the vehicle model, on the accuracy of rack force estimation. We would also like to explore the utility of our model in various EPS applications and in improving steering feel.

\bibliography{ifacconf.bib}             
 
\end{document}